\documentclass[onecolumn,english,preprintnumbers,amsmath,amssymb,floatfix]{revtex4}
\usepackage[T1]{fontenc}
\usepackage{blindtext}
\usepackage[latin9]{inputenc}
\usepackage{color}
\usepackage{array}
\usepackage{amstext}
\usepackage{graphicx}
\usepackage{esint}
\makeatletter

\@ifundefined{textcolor}{}
{%
	\definecolor{BLACK}{gray}{0}
	\definecolor{WHITE}{gray}{1}
	\definecolor{RED}{rgb}{1,0,0}
	\definecolor{GREEN}{rgb}{0,1,0}
	\definecolor{BLUE}{rgb}{0,0,1}
	\definecolor{CYAN}{cmyk}{1,0,0,0}
	\definecolor{MAGENTA}{cmyk}{0,1,0,0}
	\definecolor{YELLOW}{cmyk}{0,0,1,0}
}

\@ifundefined{definecolor}
{\usepackage{color}}{}
\@ifundefined{definecolor}
{\usepackage{color}}{}
\makeatother

\makeatother

\usepackage{babel}
\begin{document}

\title{Ground state heavy tetraquark production in heavy quark fragmentation }

\author{S.~Mohammad Moosavi Nejad$^{a,b}$}
\email{mmoosavi@yazd.ac.ir}

\author{Nahid Amiri$^a$}

\affiliation{$^{(a)}$Faculty of Physics, Yazd University, P.O. Box
89195-741, Yazd, Iran}

\affiliation{$^{(b)}$School of Particles and Accelerators,
Institute for Research in Fundamental Sciences (IPM), P.O.Box
19395-5531, Tehran, Iran}

\date{\today}

\begin{abstract}

During recent years, the study of exotic hadrons including tetraquarks and pentaquarks has attracted a lot of interests  and  more studies  are  in progress experimentally and theoretically. 
It is well-known that at sufficiently large transverse momentum the dominant production mechanism for standard heavy hadrons (mesons/baryons) is actually the fragmentation so that the same mechanism is also proposed for the production of heavy exotic hadrons. This work is the first attempt to study  the direct fragmentation of a heavy quark into ground state heavy tetraquarks in leading order of perturbative QCD. In this regard, we will present an analytical expression for the fragmentation production of  neutral hidden flavor  tetraquarks $(Qq\bar{Q}\bar{q})$ which includes most of the kinematical and dynamical properties of the process.

\end{abstract}

%\pacs{14.65.Ha, 13.88.+e, 14.40.Lb, 14.40.Nd}

\maketitle

\section{Introduction}

Since the early days of the quark model it has been assumed the possibility of the existence of exotic multiquark hadrons with the content of the valence quarks/antiquarks different from those in standard mesons and baryons. In this context, the exotic hadrons such as tetraquarks (composed of two quarks and two antiquarks) and pentaquarks (including four quarks and one antiquark) had been anticipated by Murray Gell-Mann and George Zweig in their 1964 papers \cite{GellMann:1964nj}.  The only necessary condition for these new multiquark states is to be color singlets.
However, the absence of convincing experimental evidences for such exotic structures made their investigation of marginal interest for several decades. Nevertheless, the situation dramatically changed in the last two decades when the first  explicit experimental evidence of the existence of these exotic hadrons  became available, for recent reviews see Refs.~\cite{Liu,Brambilla,Yang} and references therein.
Within the past decade several exotic hadrons, the majority pertaining to the class of heavy tetraquarks  with at least two heavy quarks, were clearly confirmed by the BELLE, BESIII and LHCb experimental collaborations. The first experimental evidence for a heavy tetraquark  was the $X(3872)$  state \cite{Belle:2003nnu} observed in 2003 by the Belle collaboration in exclusive $B^\pm \to K^\pm \pi^+ \pi^- J/\psi$ decays. The decay of this state as well as the state $Z_c (3900)^\pm$ into the $J/\psi$  meson (observed in BESIII) confirmed that these exotic particles are formed of two quarks and two antiquarks. 
Very recently, a narrow structure around 6.9~GeV, named as $X(6900)$, is observed by the CERN LHCb in the process $pp\to J/\psi J/\psi X$ \cite{Aaij}. This is the first candidate of fully-heavy tetraquarks observed in experiment which is proposed to be either the first radial (2S) excitation or the second orbital (1D) excitation of the hidden-charm  tetraquark $cc\bar{c}\bar{c}$. \\
Nowadays, convincing candidates for both the exotic tetraquark ($qq\bar{q}\bar{q}$) and pentaquark ($qqqq\bar{q}$) states are found. In the literature,  there is no definite consensus about the composition of these states and different interpretations for the tetraquark candidates are proposed, for example: molecules composed from two mesons loosely bound by the meson exchange, compact tetraquarks composed from a diquark and antidiquark bound by strong forces, hadroquarkonia composed of a heavy quarkonium embedded in a light meson, kinematic cusps, etc. The distinction between various models is a very complicated experimental task. 
Among exotic hadrons (tetraquarks, pentaquarks, hexaquarks, etc), heavy tetraquarks are of particular interest since the presence of a heavy quark increases the binding energy of the bound system and, in conclusion,  the possibility that such tetraquarks will have masses below the thresholds for decays to mesons with open heavy flavor. In this case the strong decays, which proceed through the quark and antiquark rearrangements, are kinematically forbidden and the corresponding tetraquarks can decay only weakly or electromagnetically and thus they should have a tiny decay width. 

When the strong interactions are concerned, the investigation of production mechanism of heavy standard/exotic hadrons  is a powerful tool to understand the dynamics of strong interactions.
Basically,  two different mechanisms are assumed  for the production of heavy hadrons:  recombination and fragmentation \cite{Martynenko:1995hg}. In the second mechanism, the fragmentation refers to the process of a parton which carries large transverse momentum and subsequently forms a jet containing the expected hadron \cite{Braaten:1993rw}. 
At sufficiently large transverse momentum of heavy hadron  production, the direct leading order production scheme (recombination mechanism)
is normally suppressed while the fragmentation mechanism becomes dominant, though it is formally of higher order
in the strong coupling constant $\alpha_s$ \cite{Braaten:1993rw,Kramer:2001hh}. 
In Refs.~\cite{MoosaviNejad:2020vqc,MoosaviNejad:2016qdx,MoosaviNejad:2020nsl,Nejad:2013vsa,MoosaviNejad:2014ntq,Salajegheh:2019ach,MoosaviNejad:2016scq,Salajegheh:2019nea}, we calculated the perturbative QCD  fragmentation function (FF) for a heavy quark to fragment into the $S$-wave heavy mesons and baryons at leading-order (LO) and next-to-leading order (NLO). 
In Ref.~\cite{Cheung:2004ji}, it has been pointed out that the dominant production mechanism for pentaquarks consisting of a heavy
quark is heavy quark fragmentation, similar to heavy-light mesons and baryons consisting of a heavy quark. So, this mechanism is also adopted  in our work for the production of heavy tetraquark states and we, for the first time, study the production mechanism of heavy tetraquarks in the fragmentation of a heavy quark  at lowest order of perturbative QCD.  In our calculation, tetraquarks are assumed as four-quark $Qq\bar{Q}\bar{q}$ states which are tightly bound by the color forces.  Our analytical results are presented for the fragmentation function of heavy quark $Q$ into neutral hidden flavor  tetraquarks $Qq\bar{Q}\bar{q}$ with $Q=c, b$ and $q=u, d, s$.
Having these FFs it would be possible to evaluate the production rate of these heavy states in $e^+e^-$ annihilation and hadron colliders.  In fact, the specific importance of FFs is for their model independent predictions of the cross sections at colliders. However, due to a large number of decay modes the results from $e^+e^-$ annihilation seem to be small but  sizable rates are expected in energetic hadron colliders  where a large number of them is produced.   
To assess the prospects for the experimental study of heavy tetraquarks in present and future colliders, it is important to know the accurate predictions for their production rates and, in conclusion, their FFs.

This paper is organized as follows. In Sec. II, we present the
theoretical formalism for the heavy tetraquark FF in the Suzuki model and present our numerical results, and our discussions
and conclusions are given in the last section.

\section{Heavy tetraquark fragmentation}

A motivation for studying the production of multiquark structures is for better understanding of the dynamics of strong interactions and the confinement mechanism. It is well known that the dominant mechanism to produce standard heavy hadrons  at sufficiently large transverse momentum is fragmentation so this mechanism is valid for the production of heavy tetraquarks too. Fragmentation refers to the production of a parton with a large transverse momentum which subsequently decays to form a jet containing an expected hadron. It is hence important to obtain the corresponding fragmentation function  in order to properly estimate the production rate of a specific hadronic  state.
According to the QCD factorization theorem,  the cross section for the  production of hadron $H$ in the typical 
scattering process $A+B\rightarrow H+X$, can be expressed as
\begin{eqnarray}
d\sigma=\sum_{a,b,c}\int_0^1 dx_a\int_0^1 dx_b\int_0^1 dz f_{a/A}(x_a, Q)f_{b/B}(x_b, Q) d\hat\sigma(a+b\rightarrow c+X)D_c^H(z, Q),
\end{eqnarray}
where  $\it{a}$ and $\it{b}$ are incident partons in the colliding initial hadrons $A$ and $B$, respectively,
$f_{a/A}$ and $f_{b/B}$ are the nonperturbative parton distribution function (PDFs) at the scale $Q^2$ of the partonic subprocess $a+b\rightarrow c+X$, $\it{c}$ is the
fragmenting parton  and $X$ stands for any  unobserved jet. 
Here, $D_c^H(z, Q)$ is the FF at the scale $Q$ which
can be obtained by evolving from the initial FF $D_c^H(z, \mu_0)$
using the Dokshitzer-Gribov-Lipatov-Altarelli-Parisi (DGLAP)
renormalization group equations \cite{dglap1}.
In electron-positron annihilation, one does not need
to deal with the nonperturbative  PDFs as in hadron collisions. This process has in general less contributions by background
processes compared to hadron collisions and the uncertainties introduced by parton density functions are not appeared.
The process of inclusive heavy hadron production in electron-positron annihilation, i.e., $e^+e^-\to (\gamma, Z)\to H+X$, has the cross section as \cite{Delpasand:2020vlb,Soleymaninia:2019yqz,Soleymaninia:2017xhc}
\begin{eqnarray}
\frac{d\sigma}{dx}(x,s)=\sum_{a}\int_x^1 \frac{dy}{y} \frac{d\sigma}{dy} (y, \mu_R, \mu_F) D_a^H(\frac{x}{y}, \mu_F),
\end{eqnarray}
where, $a$ stands for  the partons $a=g, u, \bar{u}, \cdots, b, \bar{b}$ and $\mu_F$ and $\mu_R$ are the factorization and renormalization scales, respectively. Generally, one can choose two different values for these scales, but a common choice consists of setting $\mu_R=\mu_F=\sqrt{s}$. Here,  $x=2E_H/\sqrt{s}$ measures the energy of hadron in units of the beam energy in the center-of-mass frame. At lowest-order of perturbative QCD, the cross section of relevant partonic subprocesses is given by \cite{Baier:1979sp}
\begin{eqnarray}\label{sigma}
\frac{d\sigma}{dy} (e^+e^-\to q\bar{q})=N_c \sigma_0 (V_{q_i}^2+A_{q_i}^2)\delta(1-y). 
\end{eqnarray}
Here, $N_c=3$ is the number of quark colors, $V_{q_i}$ and $A_{q_i}$ are the effective vector and axial-vector couplings
of quark $q_i$ to the photon and $Z$ boson. For small energies ($\sqrt{s}\ll M_Z$), for the summation of squared effective electroweak charges we have $V_{q_i}^2+A_{q_i}^2=e_{q_i}^2$ where $e_{q_i}$   is the electric charge of the quark $q_i$ \cite{Kneesch:2007ey}.
In Eq.~(\ref{sigma}), $\sigma_0=4\pi \alpha^2/(3s)$ is the leading order total cross section of $e^+e^-\to \mu^+\mu^-$ for the massless muons, in which $\alpha$ is the Sommerfeld fine-structure constant. Therefore, for the cross section of $e^+e^-\to H+X$ at LO, on has
\begin{eqnarray}\label{lo}
\frac{d\sigma^{LO}}{dx}(x,s)=N_c\sigma_0\sum_{q}e_q^2[D_q^H(x, \mu)+D_{\bar q}^H(x, \mu)]=\frac{8\pi\alpha^2}{s}\sum_{q}e_q^2 D_q^H(x, \mu),
\end{eqnarray}
where we assumed $D_q^H=D_{\bar q}^H$ for $q=u, d, s, c, b$.\\
Conventionally, the fragmentation mechanism is described by the function $D_i^H(z, \mu)$ which refers to the probability for a parton $i$ at the fragmentation  scale $\mu$ 
to fragment into a hadron $H$ carrying away a fraction $z$ of its momentum. The FFs are related to the low-energy part of the hadroproduction processes and forms the nonperturbative aspect of QCD but, fortunately, it was found that these functions for heavy hadron productions are analytically calculable by virtue of perturbative QCD (pQCD) with limited phenomenological parameters. Historically, the first theoretical effort to describe the procedure of heavy hadron production was made by Bjorken \cite{Bjorken:1977md} by using a naive quark-parton model and in the following the pQCD scheme was applied by Suzuki \cite{Suzuki:1977km}, Amiri and Ji \cite{Amiri:1986zv}. Among them, an elaborate model of fragmentation  which does contain spin information  has been proposed by Suzuki \cite{Suzuki:1985up}. This model includes most of the kinematical and dynamical properties of the splitting process and gives a detailed insight about the fragmentation process. In this approach, Suzuki calculates the fragmentation function of a heavy quark  using the convenient Feynman diagrams for the parton level of the   process and also by considering the wave function of heavy bound state which contains the nonperturbative dynamic of hadroproduction process. In fact, the Suzuki model mixes a perturbative picture with nonperturbative dynamics of fragmentation and not only predicts the
$z$-dependence of the FFs, but also their dependence on the transverse momentum of the hadron relative to the jet. 

\begin{figure}
	\begin{center}
		\includegraphics[width=0.8\linewidth]{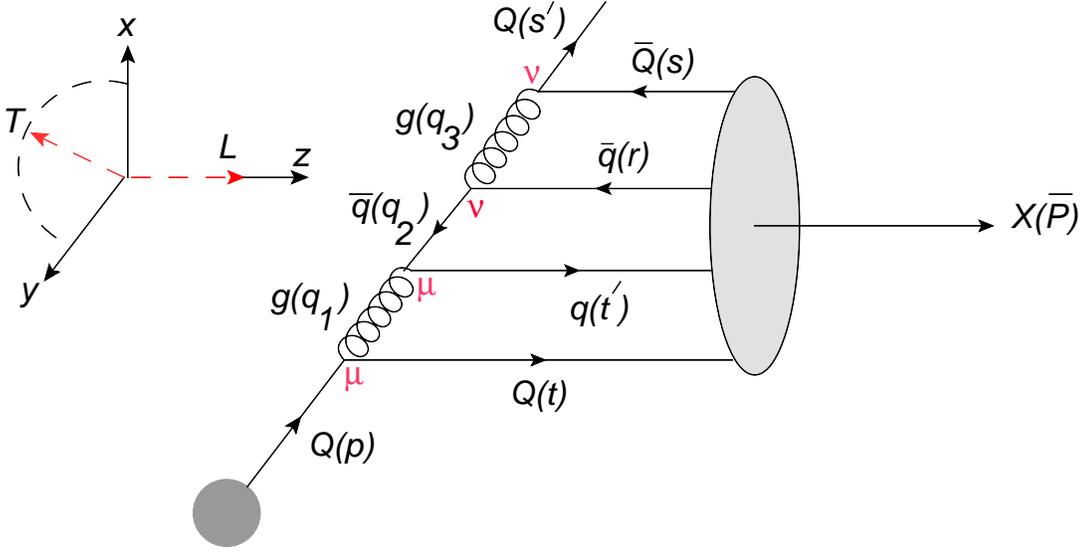}
		\caption{\label{fig1}%
			The lowest order Feynman diagram contributing to the fragmentation of a heavy quark Q into a heavy tetraquark $X(Qq\bar{Q}\bar{q})$. The four momenta are also labeled.
	}
	\end{center}
\end{figure}

Here, using the Suzuki model we focus on the heavy tetraquark FF for which the only  Feynman diagram at leading order in $\alpha_s$ is shown in Fig.~\ref{fig1}. Any other possible Feynman diagram is related to higher orders in $\alpha_s$. 
Considering this diagram we first set the relevant four-momenta  as
\begin{eqnarray}\label{kinematic}
p_\mu =[p_0, \boldsymbol{p}_T, p_L]\quad, &&\quad      s_\mu=[s_0,\boldsymbol{0}, s_L] \nonumber\\
s_\mu^\prime =[s_0^\prime, \boldsymbol{p}_T, s_L^\prime]\quad,  &&\quad     t_\mu=[t_0, \boldsymbol{0}, t_L]\\
t_\mu^\prime =[t_0^\prime, \boldsymbol{0}, t_L^\prime]\quad,  &&\quad      r_\mu=[ r_0, \boldsymbol{0}, r_L]\nonumber
\end{eqnarray}
where we assumed that the produced tetraquark $X$ moves along the $\hat{z}$-axes (fragmentation axes).  The  four-momentum of tetraquark  is also written as $\bar{P}_\mu=[\bar P_0, \boldsymbol{0}, \bar P_L]$   so that $ \bar{P}_L=s_L+r_L+t_L+t^\prime_L$. 
In the Suzuki model  the FF for the production of unpolarized ground state tetraquark is defined as \cite{Suzuki:1985up} 
\begin{eqnarray}\label{first}
D_Q^X(z, \mu_0)&=&
\frac{1}{1+2s_Q}\sum_{spin}\sum_{color}\int  |T_X|^2 \delta^3(\boldsymbol{\bar P}
+\boldsymbol{s^\prime}-\boldsymbol{p})d^3\boldsymbol{\bar P} d^3\boldsymbol{s^\prime},
\end{eqnarray}
where  $\mu_0$ is the fragmentation scale, $s_Q$ refers to the spin of the fragmenting quark and the summation
is going over the spins and colors. Four momenta are as labeled in  Fig.~\ref{fig1}. In computing Eq.~(\ref{first}), following Ref.~\cite{Suzuki:1985up}, we adopt  the infinite momentum frame where
the fragmentation  variable  in the usual light-cone form, i.e. $z=(\bar P_0+ \bar P_L)/(p_0+p_L)$, is 
reduced to a more popular one as $z=\bar P_0/p_0=E_X/E_Q$. From experimental point of view, this definition is more suitable since  refers to 
the energy fraction of the fragmenting heavy quark which is taken away by the produced tetraquark. This scaling variable ranges  as $0\le z \le 1$.\\
In Eq.~(\ref{first}), $T_X$ is
the  probability amplitude of the tetraquark   production  which, at the large momentum transfer, is expressed in terms of the  hard scattering 
amplitude $T_H$ and the process-independent distribution amplitude $\Phi_B$ as \cite{Brodsky:1985cr,Amiri:1985mm}  
\begin{eqnarray}\label{base}
T_X(\bar P, s^\prime)=\int [dx] T_H(\bar P, s^\prime, x_i) \Phi_B(x_i, Q^2),
\end{eqnarray}
where $[dx]=\Pi_{i=1}^4 dx_i\delta(1-\sum_{i=1}^4x_i)$. 
Here, $x_i$'s stand for the tetraquark energy fractions carried by its  constituent quarks which are defined as
\begin{eqnarray}\label{zare}
x_1=\frac{s_0}{\bar{P}_0}\quad,\quad x_2=\frac{r_0}{\bar{P}_0} \quad,\quad x_3=\frac{t_0^\prime}{\bar{P}_0}\quad,\quad x_4=\frac{t_0}{\bar{P}_0}\cdot
\end{eqnarray}
The energy conservation law requires that $\sum_{i=1}^4x_i=1$. 

The scheme applied to describe the probability amplitude $T_X$ as the convolution in Eq.~(\ref{base}) is convenient to absorb the soft behavior of the
bound state (tetraquark in our example) into the hard scattering amplitude $T_H$ \cite{Brodsky:1985cr}.
The short-distance part of the amplitude, i.e. $T_H$,  can be computed perturbatively from quark-gluon subprocesses at leading-order or higher-order QCD approximations.
The long-distance distribution amplitude $\Phi_B$ which contains the bound state nonperturbative dynamic of the
outgoing tetraquark, is the probability amplitude for a $Qq\bar{Q}\bar{q}$-set  to evolve
into a particular bound state.
The underlying link between hadronic phenomena in QCD at long- and short-distances is the hadronic wave function. 
In fact, the nonperturbative aspect of the hadroproduction processes is contained in the bound state of the hadron which is described by the wave function $\Phi_B$.
Following Ref.~\cite{Suzuki:1985up} and according to the Lepage-Brodsky's approach \cite{Lepage:1980fj} we 
neglect the relative motion of  the constituent quarks inside the tetraquark therefore we assume, for simplicity, that the constituent quarks are emitted collinearly with each other and  move  along the $\hat{z}$-axes.
In fact, in the Suzuki model a multiquark state  is replaced by collinear
constituents with neglecting their Fermi motion and the nonperturbative aspect of the hadroproduction
is included  in the wave function of the heavy tetraquark state. By this simplicity, the relative motion of the constituent quarks
 is effectively nonrelativistic and this allows 
one to estimate the nonrelativistic wave function of heavy tetraquark as
a delta function form, see also \cite{Nejad:2013vsa}.
Therefore, the distribution amplitude for a S-wave heavy tetraquark with neglecting the Fermi  motion reads \cite{GomshiNobary:1994eq}
\begin{eqnarray}\label{delta}
\Phi_B\approx f_B \delta(x_i-\frac{m_i}{m_X}).
\end{eqnarray}
where $m_X$ is the tetraquark  mass and  $f_B$ refers to the decay constant of hadron.
In Ref.~\cite{MoosaviNejad:2018ukp}, we studied the effect of  meson wave function on its  FF  by considering
a typical mesonic wave function which is different of the delta function 
and is the nonrelativistic limit of the solution of Bethe-Salpeter equation  with the QCD kernel \cite{Brodsky:1985cr}. There, we showed that the Fermi motion effect can be neglected with a reasonable  approximation.\\
With the delta function  approximation  for the tetraquark  wave function (\ref{delta}), we  are assuming that 
the contribution of each constituent from the tetraquark  energy is proportional
to its mass, i.e. $x_i=m_i/m_X$ where $m_X=2(m_q+m_Q)$. Considering the definitions of fragmentation variable ($z=\bar P_0/p_0$) and the fractions $x_i$'s presented in Eq.~(\ref{zare}) one also may write the parton energies in terms of the  initial heavy quark energy ($p_0$)  as: $s_0=x_1zp_0$, $r_0=x_2zp_0$, $t_0^\prime=x_3 z p_0$, $t_0=x_4 z p_0$ and $s_0^\prime=(1-z) p_0$ where $x_1=x_4=m_{Q}/m_X$ and $x_2=x_3=m_{q}/m_X$. 

In Eq.~(\ref{base}), the amplitude $T_H$ is, in essence, the partonic cross section to produce a set of heavy quarks with certain quantum numbers that in the QCD perturbation theory considering the Feynman rules  is written as
\begin{eqnarray}\label{th}
T_H=\frac{16\pi^2\alpha_s(2m_Q)\alpha_s(2m_q)m_X m_Q^2C_F}{2\sqrt{2\bar{P}_0 s_0^\prime  p_0}}\frac{\Gamma}{D_0},
\end{eqnarray}
where, $D_0=\bar P_0+s_0^\prime-p_0$ is the energy denominator, $\alpha_s(\mu)$ is the strong coupling constant at the renormalization scale $\mu$ and  $C_F$ is the color factor which is calculated using color line counting rule.   \\
In Eq.~(\ref{th}), $\Gamma$ represents an appropriate combination of the propagators and the spinorial parts of the amplitude. Considering Fig.~\ref{fig1} for the transition $Q\to X_{Qq\bar{Q}\bar{q}}$  the transition amplitude at leading order is written as 
\begin{eqnarray}
\Gamma&=&G\bar{u}(t)\gamma^\mu u(p)\bigg[\bar{u}(t^\prime)\gamma_\mu(\displaystyle{\not}q_2+m_q)\gamma_\nu v(r)\bigg]v(s)\gamma^\nu\bar{u}(s^\prime)
\end{eqnarray}
where $G$ is related to the denominator of propagators as
\begin{eqnarray}
G=\frac{1}{q_1^2(q_2^2-m_q^2)q_3^2}=\frac{1}{8(m_Q^2-p\cdot t)^2(m_Q^2+s\cdot s^\prime)^2(m_Q^2+r\cdot s+r\cdot s^\prime+s\cdot s^\prime)^2}.
\end{eqnarray}
In squaring the amplitude $\Gamma$, one needs the following  dot products of the relevant four-momenta:
\begin{eqnarray}
&& r\cdot t^\prime=m_q^2
\quad, \quad  
s\cdot t=m_Q^2
\nonumber\\
&& r\cdot s=t\cdot t^\prime=s\cdot t^\prime=m_qm_Q, 
\nonumber\\
&& p\cdot s=p\cdot t=\frac{m_Xm_Q}{2z}+\frac{zm_Q}{2m_X}(m_Q^2+p_T^2),
\nonumber\\
&& p\cdot s^\prime=\frac{1}{2}(m_Q^2+p_T^2)(1-z+\frac{1}{1-z})-p_T^2,
\\
&& r\cdot s^\prime=s^\prime\cdot t^\prime=\frac{zm_q}{2m_X(1-z)}(m_Q^2+p_T^2)+\frac{m_Xm_q(1-z)}{2z},
\nonumber\\
&&s\cdot s^\prime=t\cdot s^\prime=\frac{m_Qm_X(1-z)}{2z}+\frac{zm_Q}{2m_X(1-z)}(m_Q^2+p_T^2),
\nonumber\\
&& p\cdot t^\prime= p\cdot r=\frac{m_qm_X}{2z}+\frac{zm_q}{2m_X}(m_Q^2+p_T^2).\nonumber
\end{eqnarray}
To obtain the FF for an unpolarized S-wave heavy tetraquark  $X_{Qq\bar{Q}\bar{q}}$, considering Eqs.~(\ref{first}-\ref{th}) one has
\begin{eqnarray}\label{asli}
D_c^X(z, \mu_0) = (4\pi^2 f_B\alpha_s(2m_q)\alpha_s(2m_Q)m_Xm_Q^2C_F)^2  \int \frac{d^3\boldsymbol{\bar P}  d^3 \boldsymbol{s^\prime}\sum_{spin}|\Gamma|^2 \delta^3(\boldsymbol{\bar P}
+\boldsymbol{s^\prime}-\boldsymbol{p})}{\bar{P}_0p_0s_0^\prime (\bar{P}_0+s_0^\prime-p_0)^2}.
\end{eqnarray}
To perform the phase space integrations we start with  the
following integral
\begin{eqnarray}
&&\int \frac{d^3\boldsymbol{\bar P} \delta^3(\boldsymbol{\bar P}
	+\boldsymbol{s^\prime}-\boldsymbol{p})}{\bar{P}_0p_0(\bar{P}_0+s_0^\prime-p_0)^2}=\int  \frac{\bar{P}_0 d^3\boldsymbol{\bar P} \delta^3(\boldsymbol{\bar P}
	+\boldsymbol{s^\prime}-\boldsymbol{p})}{p_0(\bar{P}_0^2-(p_0-s_0^\prime)^2)^2}=  \frac{zp_0}{p_0(m_X^2-(p-s^\prime)^2)^2}=\frac{z}{(m_X^2-2m_Q^2+2p\cdot s^\prime)^2}.
\end{eqnarray}
According to the Suzuki approach, in Eq.~(\ref{asli}) instead of integrating over the transverse momentum of outgoing heavy quark Q, we simply replace the integration variable by its average value, i.e.
\begin{eqnarray}\label{panzdah}
\int \frac{d^3 \boldsymbol{s^\prime}}{s_0^\prime} g(z, s^\prime_T)=\int g(z, s^\prime_T)\frac{d s^\prime_L d^2 s^\prime_T}{s_0^\prime} = m_Q^2  g(z,\left\langle p_T^2\right\rangle ),
\end{eqnarray}
where the average value $\langle p_T^2 \rangle$ is a free parameter which can be specified phenomenologically.

Putting all in Eq.~(\ref{asli}), the FF of heavy tetraquark $X$ reads
\begin{eqnarray}\label{asliii}
D_Q^X(z, \mu_0)&=&N\frac{z\times \Sigma_{spin}\Gamma\bar{\Gamma}}{(m_X^2-2m_Q^2+2p\cdot s^\prime)^2}=N\frac{z\times \Sigma_{spin}\Gamma\bar{\Gamma}}{[m_X^2-(m_Q^2+\left\langle p_T^2\right\rangle)(1+z-\frac{1}{1-z})]^2}, 
\end{eqnarray}
where $N=(16\pi^2 f_B\alpha_s(2m_q)\alpha_s(2m_Q)m_qm_Q^4C_F)^2$ but it is related to the normalization condition:  $\int dz D(z,\mu_0)=1$ \cite{Suzuki:1985up,Amiri:1986zv}. Using  the traditional trace technique the  squared amplitude ($\sum|\Gamma|^2$) reads
\begin{eqnarray}
\Sigma_{spin}\Gamma\bar{\Gamma}=\frac{G^2}{z^3(1-z)}\bigg(z^6\left\langle p_T^2\right\rangle^3+\alpha z^4\left\langle p_T^2\right\rangle^2+\beta z^2\left\langle p_T^2\right\rangle-\gamma\bigg),
\end{eqnarray}
where 
\begin{eqnarray}
\alpha &=& m_Q^2(11z^2-18z+12)+6m_q^2(2z^2-z+2)+4m_qm_Q(3z^2-4z+6),\nonumber\\
\beta&=&m_Q^4(19z^4-84z^3+168z^2-144z+48)+4m_Q^2m_q^2(6z^4-37z^3+102z^2-112z+72)\nonumber\\
&&-8m_q^4(z^3-6z^2+6z-6)-16m_Qm_q^3(3z^3-12z^2+14z-12)+8m_qm_Q^3(3z^4-22z^3+54z^2-52z+24),\nonumber\\
\gamma&=&-64m_Qm_q^5(z^3+3z^2-10z+6)+8m_q^4m_Q^2(z^5-6z^4+42z^3-158z^2+236z-120)+16m_q^3m_Q^3(3z^5-\nonumber\\
&&24z^4+98z^3-196z^2+192z-80)-2m_q^2m_Q^4(6z^6-71z^5+422z^4-1232z^3+1872z^2-1424z+480)-\nonumber\\
&&4m_qm_Q^5(3z^6-40z^5+182z^4-424z^3+544z^2-352z+96)-32m_q^6(z^3-3z+2)-m_Q^6(z^2-2z+4)(3z^2-8z+4)^2.\nonumber\\
\end{eqnarray}
In the Suzuki model,  the fragmentation function depends on both the  fragmentation variable $z$ and the parameter $\left\langle p_T^2\right\rangle$ (the transverse momentum of produced hadron relative to the parent jet). In Ref.~\cite{GomshiNobary:1994eq}, it is shown that  the choice of $\left\langle p_T^2 \right\rangle=1$~GeV$^2$ is an extreme value for this quantity and any higher value will produce the peak even at lower-z regions. In this work we adopt this value for the transverse momentum of heavy tetraquark. \\
The $z$-dependence of FFs  is not yet calculable at each scale, however once they are computed at the initial scale $\mu_0$  their $\mu$ evolution can be specified by the DGLAP evolution equations  \cite{dglap1}. 
Here, we set the initial scale of fragmentation  to $\mu_0=m_X+m_Q$ which is the minimum value of the invariant mass of the fragmenting parton. Then, the FF presented in Eq.~(\ref{asliii}) should be regarded as a model for the heavy quark fragmentation $Q$ into the heavy tetraquark $X(Qq\bar{Q}\bar{q})$  at the initial scale $\mu_0$.
To present our numerical analysis, we adopt the  following input parameters \cite{Patrignani:2016xqp}:
\begin{eqnarray}
 m_c=1.67~ \textrm{GeV},\quad m_b=4.78~ \textrm{GeV},\quad f_B=0.25~ \textrm{GeV}, \quad \alpha_s(2m_c)=0.26, \quad \alpha_s(2m_b)=0.18.
\end{eqnarray}
Note that, in Ref.~\cite{Patrignani:2016xqp} two values are reported for the quark masses. The above masses  show the pole masses, i.e. the renormalized quark masses in the on-shell renormalization scheme, which correspond to the values $m_c=1.27\pm 0.02$~GeV  and $m_b=4.18\pm 0.03$~GeV for the $\overline{MS}$ masses (the renormalized quark mass in the modified minimal subtraction scheme \cite{tHooft:1973mfk}  which is intimately related to the use of dimensional regularization \cite{tHooft:1972tcz}).   The relation between the pole quark mass and the $\overline{MS}$ mass is  known to three loops  in QCD perturbation theory, details are given in Ref.~\cite{Melnikov:2000qh}. 
In our analysis, to show the effect of  variation of heavy quark masses on their corresponding FFs  we will consider the charm and bottom quark masses as   $1.27\le m_c\le 1.67$~GeV and  $4.18\le m_b\le 4.78$~GeV. 
\begin{figure}
	\begin{center}
		\includegraphics[width=0.75\linewidth]{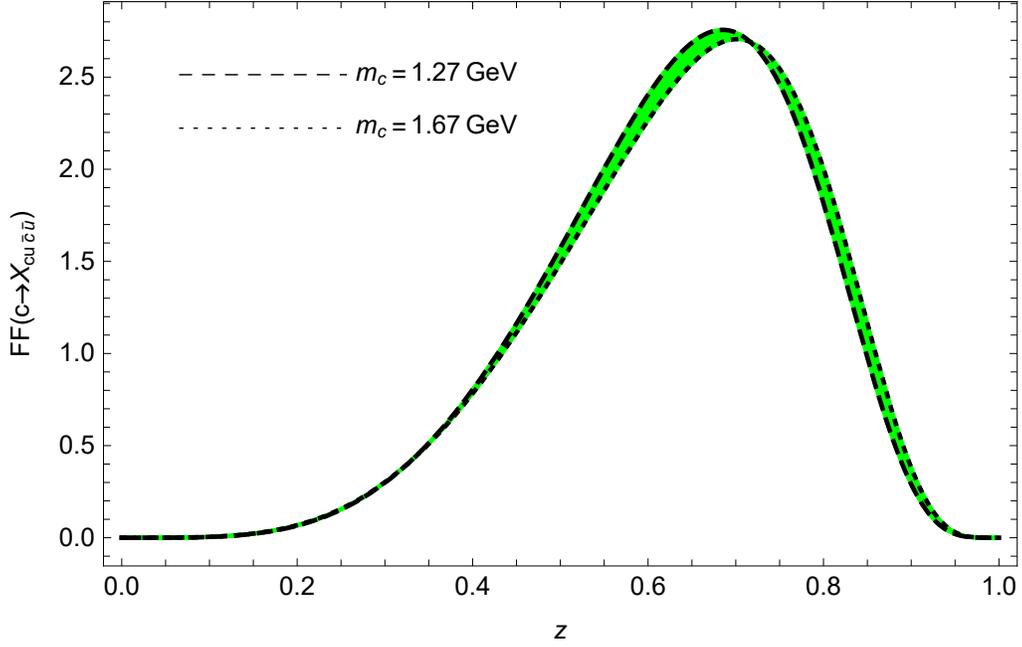}
		\caption{\label{plot1}%
			Fragmentation of  $c$-quark into  heavy tetraquark $X_{cu\bar{c}\bar{u}}$ at lowest-order of perturbative QCD.  The uncertainty band due to the variation of charm mass is also shown.}
	\end{center}
\end{figure}
\begin{figure}
	\begin{center}
		\includegraphics[width=0.75\linewidth]{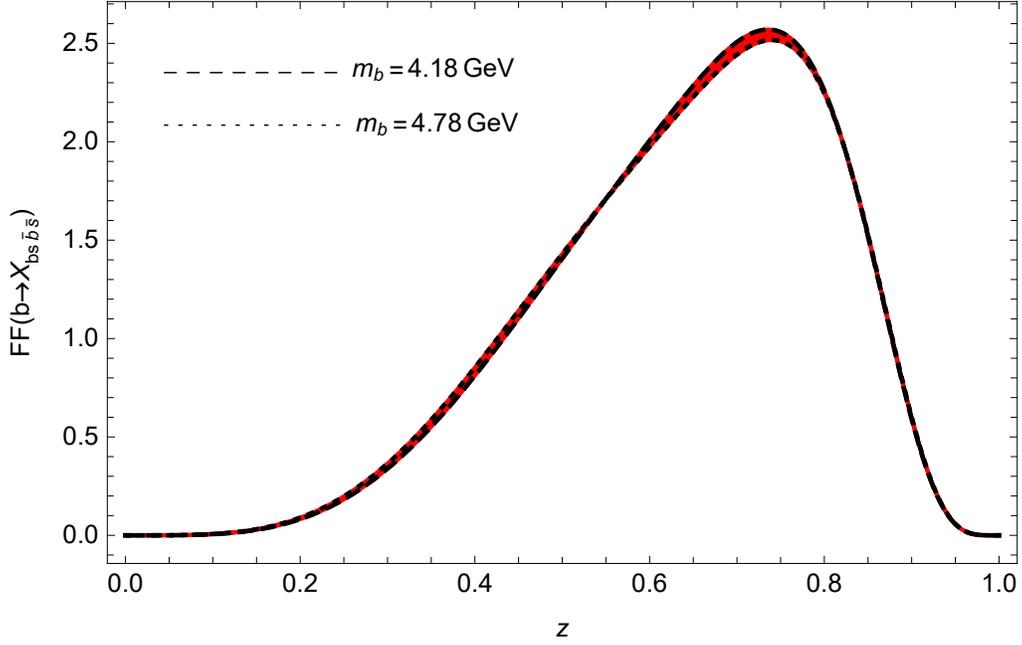}
		\caption{\label{plot2}%
			The $b\to X_{bs\bar{b}\bar{s}}$ FF at leading order perturbative QCD along with the uncertainty band of bottom mass. }
	\end{center}
\end{figure}

In Fig.~\ref{plot1}, our phenomenological prediction for the FFs of charm into the neutral hidden-charm state $X(cu\bar{c}\bar{u})$ is shown at the starting scale $\mu_0=m_c+m_X$. The uncertainty band due to the variation of charm quark mass is also plotted. Assuming $m_u\approx m_d$ this result is also valid for the transitions $c\to X(cd\bar{c}\bar{d}), X(cu\bar{c}\bar{d}), X(cd\bar{c}\bar{u})$.\\
In Fig.~\ref{plot2}, the b-quark FF into the neutral hidden-bottom tetraquark including strangeness flavor, i.e. $b\to X(bs\bar{b}\bar{s})$, is shown considering the mass uncertainty band. Since the b-quark is heavier than the c-quark, as we expected, the peak of b-quark FF shifted toward higher values of z.

Besides the heavy tetraquark FFs, their first moment is also of phenomenological interest and subject to experimental determination. It corresponds to the average fraction of energy that the tetraquark receives from the fragmenting heavy quark and is defined as \cite{Kneesch:2007ey}
\begin{eqnarray}
\left\langle z \right\rangle_Q(\mu)=\frac{1}{B_Q(\mu)}\int_0^1 zD_Q^X(z,\mu) dz,
\end{eqnarray}
where, $B_Q(\mu)$ stands for the branching fraction which is defined as  $B_Q(\mu)=\int_0^1 D_Q^X(z,\mu)dz$, where $Q=c,b$. For the tetraquark $X(cu\bar{c}\bar{u})$ the branching fraction and  the average energy fraction  are $B_c=2\times 10^{-3}$  and $\left\langle z \right\rangle_c(\mu_0)=0.635$, respectively,  and the corresponding ones for the tetraquark $X(bu\bar{b}\bar{u})$ read  $B_b=1.1\times 10^{-4}$ and $\left\langle z \right\rangle_b=0.629$. \\
Considering the inclusive differential cross section given in Eq.~(\ref{lo}), one can compute the inclusive cross section for the production of heavy tetraquark at LO as 
\begin{eqnarray}
\sigma^{LO}(e^+e^-\to X_{Qq\bar{Q}\bar{q}}+jets)=\frac{8\pi\alpha^2}{s}\sum_{q} e_q^2\int_0^1 dz  D_q^X(z, \mu),
\end{eqnarray}
By ignoring the contributions which are appeared at higher orders of perturbation, one has
\begin{eqnarray}
\sigma^{LO}(e^+e^-\to X_{Qq\bar{Q}\bar{q}}+jets)\approx \frac{8\pi\alpha^2}{s} e_Q^2 B_Q(s).
\end{eqnarray}
For example, the contributions of $c\to X_{bu\bar{b}\bar{u}}$, $b\to X_{cu\bar{c}\bar{u}}$ or $g\to X_{bu\bar{b}\bar{u}}/X_{cu\bar{c}\bar{u}}$ are related to very complicated Feynman diagrams which are in higher orders in $\alpha_s$, because they do not occur  directly. Then, their contributions are ignorable in comparison with the one occurs directly, i.e.,  $b\to X_{bu\bar{b}\bar{u}}$ and $c\to X_{cu\bar{c}\bar{u}}$  FFs (\ref{asliii}). Finally, having the branching fraction $B_Q(\mu)$  for the transition $B\to X$ at the center-of-mass energy $\mu=s$ one can compute the corresponding cross section at electron-positron annihilation.

\section{Conclusion}

Besides the standard mesons and baryons there exist hadrons made of more than three quarks/antiquarks or valence gluons. These new types of hadrons consist of exotic states including glueballs, hybrids, tetraquarks, pentaquarks, hexaquarks, etc. In last two decades a large number of these exotic states, especially tetraquarks and pentaquarks, have been observed in the particle factories. However, some of them are now well-established there are also some doubts on existence of some other members. Therefore, exact determinations of the nature, structure and quantum numbers of these exotic states need more experimental efforts. Specifically, there are many exotic states which are proposed in theory but their existence need to be confirmed by the experiments.

It should be noted that, most of heavy multiquark particles discovered in the experiment have hidden-charm or hidden-bottom quark structure, i.e., they contain $c\bar{c}$ or $b\bar{b}$ in their inner structures. The study of  their production mechanism and using it as a probe to the structure of  
hadrons are among the most active research fields in particle and nuclear physics.
In fact, the challenge is to understand
the nonperturbative transition from high energy $e^+e^-$, photon-hadron, and hadron-hadron collisions to physical exotic states.\\
It is well-known that the dominant mechanism to produce heavy hadrons at high transverse momentum is fragmentation; the production of a high energy parton followed by its fragmentation into the heavy bound states. 
In this work, using the Suzuki model we, for the first time,  studied the fragmentation function of a heavy quark into an unpolarized S-wave heavy tetraquark at leading order perturbative QCD. To be specific, we analytically computed the FF of bottom and charm quark into the neutral hidden flavor tetraquarks ($X|Qq\bar{Q}\bar{q}>$) at the initial scale $\mu_0=m_X+m_Q$.  Using the extracted FFs one can compute  the production rates of heavy tetraquarks in hadron colliders.\\ 
To have more accurate FFs, one can think of other effects such as the Fermi motion of constituents, the tetraquark mass effect, NLO radiative corrections, etc. Although, including the effects such as the Fermi motion of constituents leads to very complicated integrals which should be solved numerically \cite{MoosaviNejad:2018ukp}.

\end{document}